# ANOMALOUS ATTENUATION OF EXTRAORDINARY WAVES IN THE IONOSPHERE HEATING EXPERIMENTS

N.A. Zabotin[1], A.G. Bronin[1], V.L. Frolov[2], G.P. Komrakov[2], N.A. Mityakov[2], E.N. Sergeev[2], G.A. Zhbankov[1]

[1] *Rostov State University, Rostov-on-Don, Russia*

[2] *NIRFI, Nizhni Novgorod, Russia*

Abstract. Multiple scattering of radio waves by artificial random irregularities HF-induced in the ionosphere F region may cause significant attenuation of both ordinary and extraordinary waves together with common anomalous absorption of ordinary waves due to their non-linear conversion into plasma waves. To demonstrate existence and strength of this effect, direct measurements of attenuation of both powerful pump wave and weak probing waves of extraordinary polarization have been carried out during an experimental campaign on September 6, 7 and 9, 1999 at the Sura heating facility. The attenuation magnitude of extraordinary waves reaches of 1–10 dB over a background attenuation caused by natural irregularities. It is interpreted in the paper on the base of the theory of multiple scattering from the artificial random irregularities with characteristic scale lengths of 0.1–1 km. Simple procedure for determining of irregularity spectrum parameters from the measured attenuation of extraordinary waves has been implemented and some conclusions about the artificial irregularity formation have been obtained.

## *Introduction*

Attenuation of radio waves reflected in the heated volume of the ionosphere is observed in ionospheric modification experiments since early 1970's [Allen *et al.,* 1974; Belikovich *et al.,* 1975; Berezin *et al.,* 1987; Frolov *et al.,* 1997]. Main attention was given to anomalous attenuation of the ordinary wave, known also as 'anomalous absorption' (AA). The AA is based on the well known mechanism of ordinary wave conversion into the upper hybrid plasma waves via scattering from small-scale (1-50 m) irregularities. The theory of this mechanism has been well developed (see, for example, statement of classical approaches in [Vas'kov and Gurevich, 1975; Grach *et al.,* 1978; Das and Fejer, 1979; Robinson, 1989] and some new results in [Bronin *et al.*, 1999]).

Effect of anomalous attenuation for extraordinary wave have been studied rather poorly in comparison with ordinary wave. It was expected that it is negligible in most cases since the conversion mechanism does not work for extraordinary waves in the ionosphere. However, there are some experimental results showing that anomalous attenuation for the extraordinary wave may reach values comparable with the ordinary wave AA (see, for example, [Erukhimov *et al*, 1980]). Below in the paper we will discuss experimental data obtained during purposeful experimental campaign at the Sura heating facility (Vasil'sursk, Russia) performed in September, 1999, which have clearly demonstrated permanent existence of the anomalous attenuation for both pump and probing extraordinary waves.

The only actual mechanism of the anomalous attenuation of the extraordinary wave is the small-angle multiple scattering from intermediate-scale (transversal with relation to the geomagnetic field scale length $l_\perp \sim 0.1$-1 km) electron density irregularities. As it has been recently established (see [Zabotin *et al,* 1998] and references therein), multiple scattering under conditions of strong ionospheric refraction results in the spatial redistribution of radiation reflected from the ionosphere with considerable decrease in its intensity in a vicinity of the sounding station as compared with radiation distribution in absence of the irregularities. This attenuation, perceived by a ground-based observer as anomalous (non-collisional) attenuation of both ordinary and extraordinary waves, may be so large as 10 dB or more.

Thus, when intermediate-scale and small-scale irregularities are developed together, multiple scattering may cause the attenuation of extraordinary wave and contribute into the attenuation of the ordinary wave together with 'anomalous absorption'. The purpose of the present paper is to study in detail the role of the attenuation of extraordinary waves due to scattering mechanism and demonstrate possibility to use such X-mode probing for diagnostics of artificial irregularities in the ionospheric plasma.

## *2. Theoretical grounds for anomalous attenuation effect*

The AA of the ordinary wave is closely connected with generation of artificial small-scale ($l_\perp < 50$ m [Frolov *et al.,* 1997]) irregularities. These irregularities are produced in plasma by the thermal parametric instability of the powerful ordinary wave [Vas'kov and Gurevich, 1975; Grach *et al.,* 1977; Das and Fejer, 1979]. Conversion of the pump wave into the upper-hybrid plasma waves is accompanied by growth of the irregularity amplitude what results in fast development of the AA phenomenon (several seconds after the pump wave is turned on). When they aroused, the small-scale irregularities are equally able to influence both powerful pump wave (PW) and probing wave amplitudes. For a known spatial spectrum of artificial irregularities the AA value may be most conveniently estimated using the conversion cross-section [Bronin *et al.,* 1999a].

The AA affects only the ordinary wave because extraordinary one is reflected below the upper-hybrid coupling region. On the contrary, the usual (without the conversion) scattering from the intermediate-scale irregularities is able to affect both O- and X-mode waves. It should be noted that such irregularities are often present under natural ionospheric conditions [see, e.g., Szuszczciewicz, 1987]. They are also generated in heating experiments due to, for example, the self-focusing instability of the PW [Vas'kov and Gurevich, 1976]. Correspondingly, measurements of the radio wave attenuation in the undisturbed ionosphere [Setty *et al.,* 1971; Vodolazkin *et al.,* 1983; Bronin *et al.,* 1999b] demonstrate existence of the anomalous attenuation which cannot be explained by collisional losses of the radio waves. Note that small-scale irregularities ($l_\perp <$ 50 m) are suppressed in the undisturbed ionosphere what manifests itself in power-law behavior (with negative index) of the natural irregularity spatial spectrum [Szuszczciewicz, 1987], so the mode conversion mechanism also cannot explain significant attenuation of the sounding signal. Only common multiple scattering may provide the observed magnitude of the anomalous attenuation of radio waves in the natural ionosphere.

Theoretical description of the vertical sounding signal multiple scattering from the ionospheric irregularities is a fairly complex task. In a general case it is necessary to take into account the regular refraction and the hyrotropy of the medium. If the intensity of the reflected signal is of interest only, the task may be solved by means of the radiative transfer theory [Bronin and Zabonin, 1992; Zabotin, 1993; Zabotin *et al.*, 1998; Bronin *et al.*, 1999c]. For the model of the randomly irregular plane ionospheric layer this approach allows one to obtain an approximate solution in analytical form. According to [Zabotin *et al.*, 1998] the energy flux at a point with coordinate $\vec{\rho}$ on the Earth's surface may be written as:

$$\tilde{P}(\vec{\rho}) = \tilde{P}_0\left[\vec{\rho} + \vec{D}(\theta_1, \varphi_1)\right] \cdot \left| \frac{\partial(\rho_{0x}, \rho_{0y})}{\partial(\theta, \varphi)} \right| \cdot \left| \frac{\partial(\rho_{0x} - D_x, \rho_{0y} - D_y)}{\partial(\theta, \varphi)} \right|^{-1}_{\theta=\theta_1, \varphi=\varphi_1} . \qquad (1)$$

Here $\tilde{P}_0$ is the flux of radiation energy at the point $\vec{\rho}$ in absence of the scattering; $\theta_1$ and $\varphi_1$ are effective angles of arrival of the energy flux, determined by the transcendent equation

$$\vec{\rho} - \vec{\rho}_0(\theta_1, \varphi_1) + \vec{D}(\theta_1, \varphi_1) = 0 \ , \qquad (2)$$

where $\vec{\rho}_0(\theta, \varphi)$ is the point of arrival at the Earth's surface of the ray which angles of arrival in absence of the irregularities are $\theta, \varphi$, the quantity $\vec{D}$ depends on both geometry of the ray paths in the plane layer and the spatial spectrum of the irregularities.

According to expressions (1) and (2), an observer situated at the point $\vec{\rho}$, will detect two effects caused by the multiple scattering: the change of arrival angles and the decrease or increase in the intensity of the received wave after its reflection from the ionosphere. For the vertical sounding of the quiet mid-latitude ionosphere the anomalous attenuation, defined as $L = 10 \cdot \lg\left(\tilde{P}_0 / \tilde{P}\right)$, may reach of 10 dB or more [Zabotin *et al.*, 1998].

Expression (1), together with parameters of the ionospheric layer (electron density profile, etc.) and parameters of the spatial spectrum of irregularities specified, forms the quantitative model of the anomalous attenuation. This model can be used to solve the inverse problem of the irregularity diagnostics: the spectrum parameters are determined by fitting the experimental data in spirit of the least squares method. Using the extraordinary wave attenuation measurements one may estimate the parameters of the intermediate-scale irregularity spectrum, then separate contributions from the multiple scattering and from the mode conversion mechanism into the anomalous attenuation of the ordinary wave and finally determine the spectrum properties for the small-scale irregularities.

Full implementation of this plan, however, requires extended experimental data set that currently cannot be provided by the experimental facility being at our disposal. For example, simultaneous measurements of attenuation of the ordinary and extraordinary waves using close frequency grid are desirable but not carried out now. It would be nice also to know the absolute value of the attenuation what means, in particular, that the "background" attenuation, presumably caused by the natural irregularities should be determined before heating of the ionospheric plasma. Present technique used for the anomalous attenuation observations is based on the relative measurement principle.

However, it will be shown below that valuable results may be derived from experimental data obtained with the available technique.

## *2. Experimental results*

The experiments with the Sura heating facility were carried out on September 6, 7 and 9, 1999 in the evening or night hours when collisional absorption in the ionosphere D and E regions could be considered negligible. Heating was provided by coherent work of three 250 kW transmitters. With account of the antenna gain the effective radiated power (ERP) was 150-250 MW depending on the PW frequency (150 MW ERP for 5.75 MHz and 250 MW ERP for 7.8 MHz). In the September 9 session the heater power was varied but we use for the purposes of our present work only data of homogeneous subset of the heating cycles corresponding to 150 MW ERP. The duration of each heating cycle on September 6 and 7 was 15 minutes including the active phase of 5 minutes length (heater on) and off-time of 10 minutes. On September 9 another scheme was used: 3 min on — 7 min off. Overall duration of the measurement series included into processing was 3.9 hours (15 heating cycles) in the first case, 3 hours (12 heating cycles) in the second case, and 4h50m in the third case (12 heating cycles). In the experiment on September 6 the heating was performed by the high-power HF wave of ordinary polarization at the pump frequency 5,752 MHz. Amplitudes of probing waves were simultaneously recorded at 7 frequencies: 4.069, 4.669, 5.669, 6.069, 6.269, 6.424 and 6.849 MHz. In the experiment on September 7 the extraordinary PW at the frequency 7.815 MHz was used. Simultaneous amplitude measurements were carried out with 6 probing waves at the following frequencies: 5.424, 6.624, 7.224, 7.624, 7.789 and 8.024 MHz. On September 9 the used frequencies were: 5.752 MHz for the PW and 4.469, 4.969, 5.369, 5.569, 5.769, 5.969, 6.169 MHz for the probing waves. A 'Katran' receiver (with pass band of 4 kHz) was used for registering of the PW amplitude and 'Brusnika' receivers (with pass band of 1 kHz) were used for registering of the probing wave amplitudes. In all series only probing waves of extraordinary polarization were used. The probing wave transmitter emitted pulse signals of the linear polarization with duration of 100 microseconds. After reflection from the ionospheric layer these signals were received by the antenna of the extraordinary polarization with 15 dB suppression of the ordinary component.

During the heating experiment the electron density profile was controlled by both vertical and oblique radio sounding. According to these data, the F2 layer critical frequencies exceeded as a rule both PW frequency and probing wave ones. The only exclusion took place at the end of the September 7 series, so the data from 3 last cycles (7 cycles for the top frequency probing wave) were to be excluded from the processing.

An example of the initial data (records of the amplitude for a probing wave and the PW for the September 6 session) is shown in Figure 1. Interval between the data points is 0.2 sec. The effect constituting the subject of this paper manifests itself distinctly already in data of such kind: It is seen that periods of the heater turn-on are accompanied by significant drop of the extraordinary probing wave amplitude. To determine the details of this process it is necessary to weaken the natural variability of the signal. It is achieved by averaging of the original data over a number of heating cycles during the same experimental session. Instants of switching the pump wave on (off) serve as the temporal reference points for this procedure. The result of such averaging for experimental data obtained on September 6, 7 and 9 is shown in Figures 2-4 respectively.

Four successive stages of the probing signal evolution can be discriminated within a heating cycle. The first stage, immediately preceding the heater turn-on, is characterized by random variations of the signal amplitude near some median value $A_{\mathrm{off}}$. Right away the PW turn-on the transitional stage follows during which the mean signal amplitude decreases gradually to smaller value $A_{\mathrm{on}}$. This is the development stage of the anomalous attenuation effect. The next stage lasts till the PW turn-off. It is characterized by random amplitude variations near the $A_{\mathrm{on}}$. After the PW turn-off the relaxation stage occurs during which the ionosphere restores its natural undisturbed state and the signal amplitude increases gradually to the initial median value $A_{\mathrm{off}}$. Note that in a varying degree these features of the probing signal behavior are characteristic of all used frequencies for all three observation days. Features of the reflected PW amplitude are also of interest. There is nothing special in the attenuation of the ordinary PW for the data obtained on September 6: This effect is attributed mainly to the mode conversion mechanism and is well studied before [Belikovich *et al.*,1975; Robinson, 1989]. In the present case the attenuation was about 11 dB and development time for this effect was of order of one second, in accordance with classical notions about the thermal parametric instability. But anomalous attenuation of a significant magnitude occurs also for the extraordinary PW used during two other observational sessions, on September 7 and 9. It was near 12 dB with development time about 40 sec on September 7 and about 4-5 dB with development time near 20 sec on September 9. We can thus conclude that such anomalous attenuation of extraordinary waves (both probing and pump) is a quite common effect in the heating experiments.

There are some additional common points of the dependencies presented in Figs. 2-4 that in fact are not connected with the heating activity. For example, in Fig. 2 near 30, 350 and 380 sec one can see sharp changes of the probing wave amplitude simultaneous for all used frequencies. This is a result of the nearby ionosonde activity which periodicity coincided with the heater cycling. We have excluded these time intervals from the data processing.

We will characterize the anomalous attenuation effect by the relative weakening of the wave intensity expressed in decibels in the following way: $L_R = 20 \cdot \lg(A_{\rm off}/A_{\rm on})$. Two other parameters of interest are the development time $\tau_{\rm dev}$ and the relaxation time $\tau_{\rm rel}$ defined as the time intervals during which the decrease (increase) of the wave mean amplitude achieves of the $1/e \approx 0.37$ of its full magnitude $A_{\rm off} - A_{\rm on}$. Both of them can be easily determined by application of the following fits to the averaged data within corresponding temporal interval: $A(t) = A_{\rm on} + (A_{\rm off} - A_{\rm on}) e^{-t/\tau_{\rm dev}}$ and $A(t) = A_{\rm on} + (A_{\rm off} - A_{\rm on})\left(1 - e^{-t/\tau_{\rm rel}}\right)$.

Results of data processing are presented in Figs. 5-7. Maximal relative attenuation of probing waves is about of 9.6 dB, the minimal registered one is about of 1.5 dB (see Fig. 5). All three data sets show common tendency of gradual decrease in the relative attenuation $L_R$ when the sounding frequency is increased. This distinction of the present results from those presented in [Erukhimov *et al*, 1980] is caused evidently by the difference in the PW ERP (20 MW ERP in [Erukhimov *et al*, 1980] and 150-250 MW ERP in our work). More powerful HF exposure leads to creation of stronger irregularities at larger distances from the heating region. So in this case the anomalous attenuation effect manifests itself stronger at the lower frequencies.

Two characteristic times, $\tau_{\rm dev}$ and $\tau_{\rm rel}$, do not demonstrate clear regular change with the frequency growth. One should notice, however, that maximal development time corresponds to reflection from the heating region altitudes in each of three sequences. For the most part, the development time values lie in the interval 2-40 sec corresponding nicely to known time of development of artificial intermediate-scale irregularities [Frolov *et al.*, 2000; Frolov *et al.*, 1997]. Note, however, that on September 9 the $\tau_{\rm dev}$ was steadily kept near the smallest values 2-5 sec. Values of characteristic times corresponding to three highest frequencies on September 7 are visibly aside from the common consistent pattern. It can be explained by closeness of these frequencies to the F layer critical frequency (~7.9 MHz). This relates also to the relaxation time values for the same frequencies. Excluding them from consideration one can infer that observed spreading of the relaxation time values is very small for a given frequency, for all three series they lie in the interval 13-70 sec. The latter interval overlaps essentially with that one for the development time, though one can notice that the $\tau_{\rm rel}$ values are as a rule higher than the $\tau_{\rm dev}$ ones. These characteristic times properties are explained naturally on the assumption that the process primarily responsible for both irregularity formation and decay out of the heating region is universal: It is the ambipolar diffusion of the electron density disturbance emerged inside the heating region along the magnetic field lines. But on the development stage this diffusion is enforced by the thermal pressure of the powerful HF field, whereas relaxation takes place when this driving force is switched off.

There are not any marked differences between the considered effects caused by ordinary or extraordinary pump waves. It is connected probably with the circumstance that probing waves, frequencies for which would allow them to reflect inside the heating region, are practically absent in this data set. So effects outside of the heating region have been studied here.

The observed attenuation of the extraordinary waves (the probing waves in all sessions of measurements as well as the PW in two observational days) is characterized by large values (the amplitude drops by factor of 1.5-4). The only known mechanism able to explain it has been briefly described in the previous section: this is the multiple scattering from irregularities with scale lengths 0.1-1 km. The relations provided by the theory can be used to determine the irregularity characteristics from the anomalous attenuation data.

## 3. Inverse problem

The theoretical relations cited in Section 1 may be used to obtain information about the spatial spectrum of artificial intermediate-scale irregularities from the anomalous attenuation data. Solving of some kind of the inverse problem is required for that.

Following to the standard conceptions [e.g., Eruchimov *et al.*, 1987] we shall believe that artificial irregularities in the scale length range of interest are strongly elongated along the geomagnetic field lines and characterize them by the following spatial spectrum:

$$F(\vec{\varkappa}) \propto \left(1+\varkappa_{\perp}^{2}/\varkappa_{0\perp}^{2}\right)^{-\nu/2} \delta\left(\varkappa_{\parallel}\right), \tag{3}$$

where $\varkappa_{\perp}$ is the component of the irregularity harmonic $\vec{\varkappa}$ orthogonal to the geomagnetic field lines, $\varkappa_{0\perp} = 2\pi/L_m$, $L_m$ is the irregularity upper scale length, $\delta(x)$ is the Dirac delta. It is convenient to normalize the spectrum on the structure function value $D_N\left(\vec{R}\right)=\left\langle\left[\frac{\Delta N}{N}\left(\vec{r}+\vec{R}\right)-\frac{\Delta N}{N}\left(\vec{r}\right)\right]^{2}\right\rangle \equiv \delta_R^2$ taken for the scale length $R = l_{\perp} = 1$ km. Thus the spectrum has three parameters: $\delta_R$, $\nu$ and $L_m$.

The information available in our data sets is not sufficient to determine all these spectrum parameters simultaneously (mainly due to small number of the probing wave frequencies). That is why we assign to two of them rather typical predetermined values, $L_m = 10$ km and $\nu = 2.5$, proposing to look for dependence of $\delta_R$ on the altitude.

Note that our data on the relative anomalous attenuation in principle do not contain any information on the background level of the attenuation caused by natural ionospheric irregularities. So we need to predetermine also the model of natural irregularities. Let us suppose that natural irregularities are described by the same spectrum (4) with the same parameters $L_m = 10$ km and $\nu = 2.5$, and their amplitude (we shall denote it as $\delta_{\rm N}$ to distinguish it from the current value $\delta_R$ which is changed under the heating influence) does not depend upon the altitude. In other words, we suppose that heating of the ionosphere results only in change of the irregularity amplitude altitude distribution, while the spectrum slope remains unchanged. In this case the quantity $\delta_{\rm N}$ is a free parameter in our calculations.

The above model of the spectrum is not entirely adequate neither for artificial irregularities nor for natural ones. The real spectra are known to demonstrate somewhat different behavior for different ranges of scale lengths [Szuszczciewicz, 1987; Erukhimov *et al.*, 1987]. For example, they may have a hump near 0.3-0.5 km [Erukhimov *et al.*, 1987]. We do not provide in this paper an ultimate diagnostic procedure taking into account numerous properties of real irregularity spectra, electron density profile, etc. This Section results are to demonstrate only possibility of such diagnostics.

The general scheme for calculation of the irregularity amplitude looks as follows (see Fig. 8). At first one should determine for each frequency of the probing signal the background attenuation $L_{\rm N}$ corresponding to the given value of $\delta_{\rm N}$. Total attenuation after completion of the effect development is $L = L_{\rm N} + L_R$, where $L_R$ is the experimentally determined quantity as stated in the previous Section. The resulting value of $\delta_R$ can be determined by solving the equation $L = \tilde{P}(\vec{0})$ where $\tilde{P}(\vec{0})$ is given by the expression (1) which is a non-linear function of $\delta_R$. The non-linear equation $L = \tilde{P}(\delta_R)$ can be solved numerically by a simple (e.g., dichotomy) method .

We make two simplifications in our calculations. The real profile of the electron density altitude dependence is substituted by the linear profile with the same slope at the reflection point. It is possible owing to a key role in the multiple scattering process of the irregularities located near the reflection level. Also the approximation of isotropic ionospheric plasma was used in calculation of quantity $\tilde{P}$.

Dimensionless quantity $\Delta N/N$ depends not only on electron density perturbation $\Delta N$, but also on the mean value $N$. The latter often changes significantly in the considered range of altitudes. On the other hand, one would expect that rapid ambipolar diffusion along the geomagnetic field lines tries to equalize just the quantity $\Delta N$, if it is generated inside the heated region under influence of the powerful high-frequency field. Thus the quantity $\Delta N$ may give more valuable physical information namely for the artificial irregularities. That is why we present the results of the inverse problem solving in the form of altitude dependencies of $\Delta N$ for different values of $\delta_{\rm N}$ (Fig. 9).

Analysis of the Fig. 9 shows first of all that artificial irregularities are created in the broad altitude range (more than 40 km — the value determined by the used probing wave frequency band) which significantly exceeds the heating region vertical dimension (~1-2 km). No doubt that broadening of the used frequency band would allow one to detect artificial irregularities also at other altitudes. Effect of the supposed $\delta_N$ value growth manifests itself in the increase of the resulting irregularity $\Delta N$ which is more quick for larger altitudes. Zero $\delta_N$ value results in the expressed general tendency for the $\Delta N$ to decrease with the altitude growth which is difficult to explain. But already for relatively low and quite realistic value of the natural kilometer irregularity amplitude $\delta_N \sim 0.001$ we have a rather reasonable altitude dependence of the $\Delta N$, when this quantity is approximately constant in a wide range of altitudes. Relative decrease in the $\Delta N$ magnitude observed in these dependencies near the heating region altitudes may be explained by the pressure gradient caused by increased temperature of particles and by action of powerful high-frequency field [Gurevich, 1978].

## ***Conclusion***

An important result of the September 1999 heating campaign at the Sura facility is clear demonstration of permanent occurrence of considerable (1.5-12 dB) anomalous attenuation of both pump and probing extraordinary waves. This attenuation cannot be explained by the mode conversion mechanism when scattering from the small-scale ($l_\perp < 50$ m) irregularities which works only with ordinary waves. The mechanism of strong attenuation of waves independent on their polarization has been suggested basing on the theory of the multiple scattering from irregularities with characteristic scale lengths $l_\perp \sim 100$–$1000$ m. The experimental results discussed in the present paper can be considered as independent confirmation of the conclusions of this theory on an important role of the multiple scattering in the ionosphere both for natural and artificial conditions.

Very high power (150-250 MW ERP) of the pump wave in the present experimental campaign might create conditions for some complication of the studied effect due to possible development of large-scale (larger than 10 km) irregularities inside the heating region. Multy-beam reflection from the ionospheric layer with the large-scale irregularities is accompanied by the increase of the averaged signal amplitude [Zabotin and Zhbankov, 2000]. Manifestation of this effect is well seen in the pump wave amplitude behavior in Fig. 2: After arrival at the minimal value the amplitude is then slightly increased. Experimentally it has been established earlier that this effect occurs for the heating effective radiated power larger than 20 MW [Berezin *et al.*, 1987]. Thus in the considered data this effect can coexist with the anomalous attenuation weakening the latter in some degree. Therefore, carrying out and analyzing the results of special experimental campaign for the low heating power, when the anomalous attenuation effect could be observed in its classical form, would be of particular interest.

Basing on anomalous attenuation data a simple algorithm of the inverse problem solving has been implemented in this work to determine the distribution of the HF-induced intermediate-scale electron density disturbances $\Delta N$ in a wide range of altitudes. The calculations give reasonable values of $\Delta N$, which are in good agreement with modern concepts of artificial irregularity formation. Note that this is the direct opportunity to measure the $\Delta N$ distribution along the geomagnetic field line. The altitude range under consideration is limited only by the frequency band used for the probing wave diagnostics.

Of course, the method may be improved. Measurements of the anomalous attenuation of the extraordinary and ordinary waves simultaneously allow one, in principle, diagnostics of artificial irregularities with very different scales: small-scale irregularities (~1–50 m) and intermediate-scale irregularities (~100–1000 m). An advanced technique should provide also possibility to measure the background anomalous attenuation caused by the natural irregularities.

*Acknowledgements.* The work was carried out under support of Russian Foundation for Basic Research (grants No.99-02-17525 and No.99-02-16479).

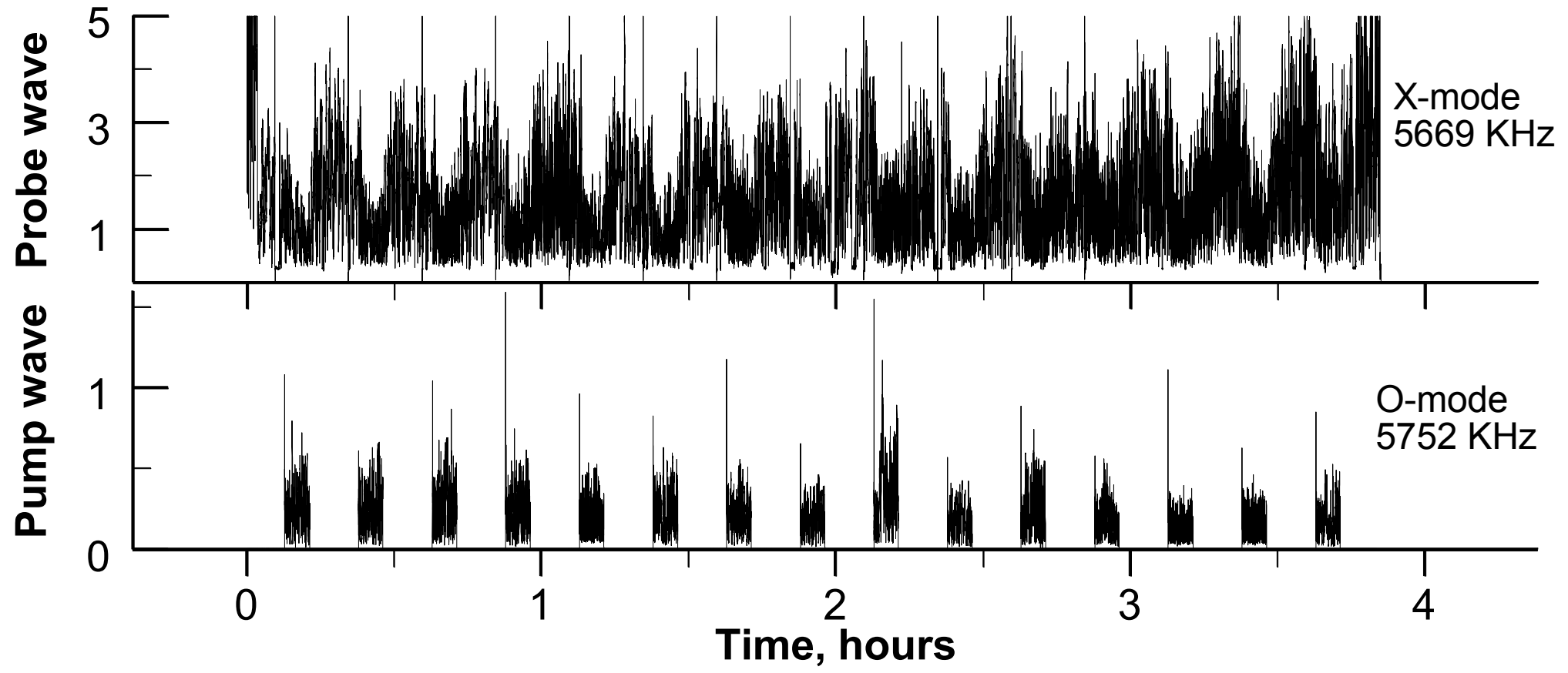

Figure 1.

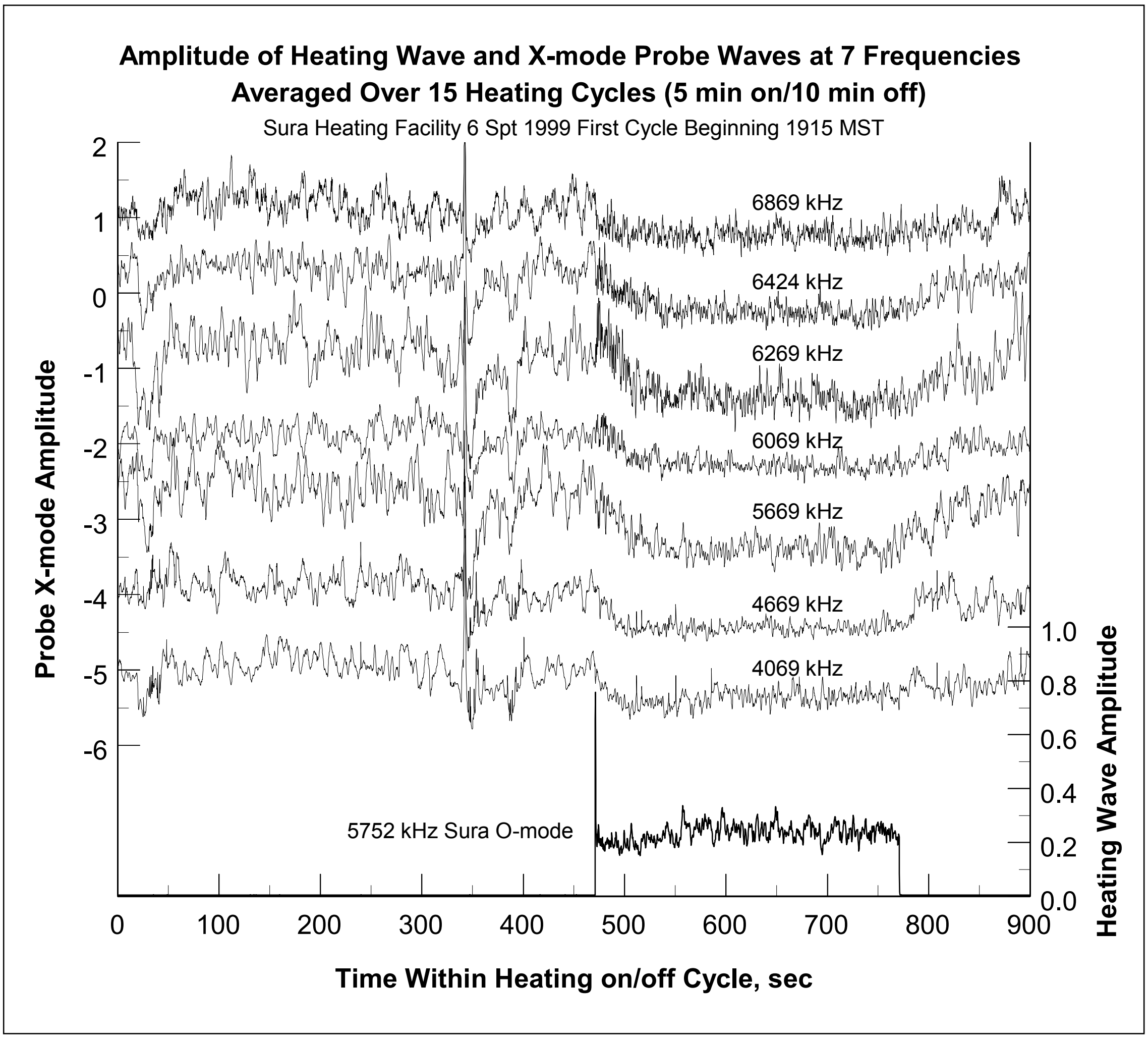

Figure 2.

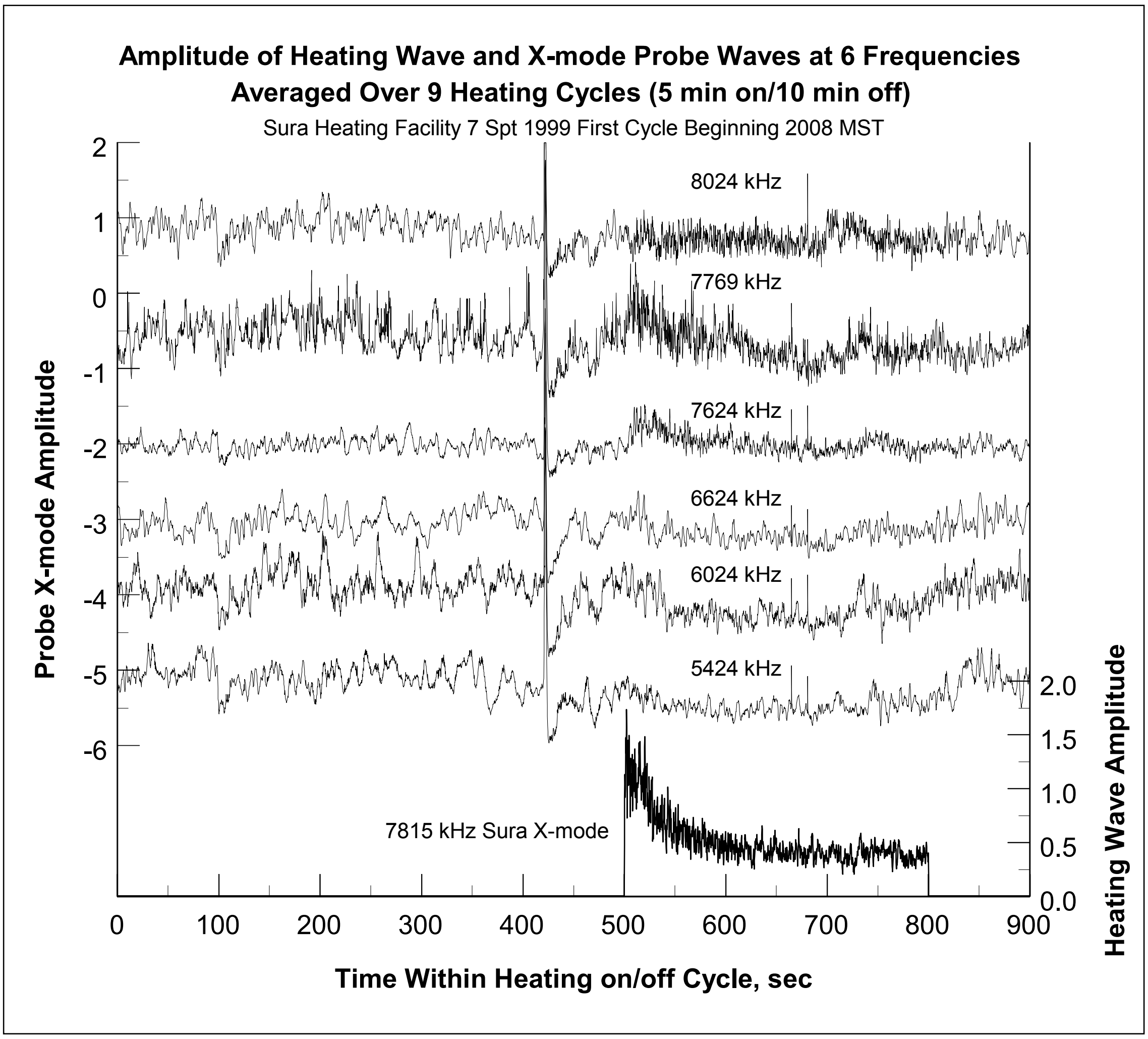
Amplitude of Heating Wave and X-mode Probe Waves at 6 Frequencies
Averaged Over 9 Heating Cycles (5 min on/10 min off)
Sura Heating Facility 7 Spt 1999 First Cycle Beginning 2008 MST
8024 kHz
7769 kHz
7624 kHz
6624 kHz
6024 kHz
5424 kHz
7815 kHz Sura X-mode
Probe X-mode Amplitude
Heating Wave Amplitude
Time Within Heating on/off Cycle, sec
2
1
0
-1
-2
-3
-4
-5
-6
2.0
1.5
1.0
0.5
0.0
0
100
200
300
400
500
600
700
800
900

Figure 3.

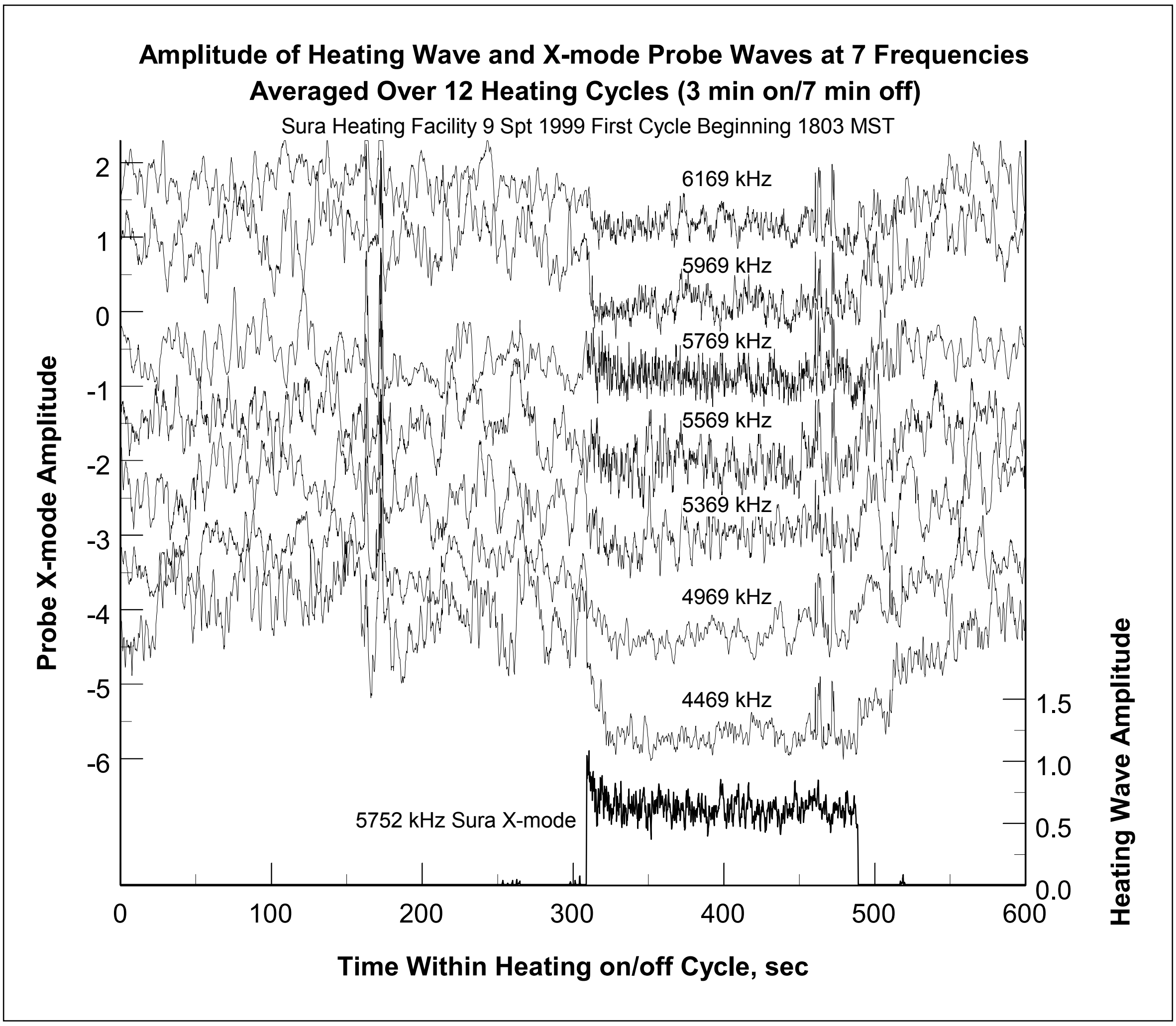
Amplitude of Heating Wave and X-mode Probe Waves at 7 Frequencies
Averaged Over 12 Heating Cycles (3 min on/7 min off)
Sura Heating Facility 9 Spt 1999 First Cycle Beginning 1803 MST
6169 kHz
5969 kHz
5769 kHz
5569 kHz
5369 kHz
4969 kHz
4469 kHz
5752 kHz Sura X-mode
Probe X-mode Amplitude
2
1
0
-1
-2
-3
-4
-5
-6
Heating Wave Amplitude
1.5
1.0
0.5
0.0
0
100
200
300
400
500
600
Time Within Heating on/off Cycle, sec

Figure 4.

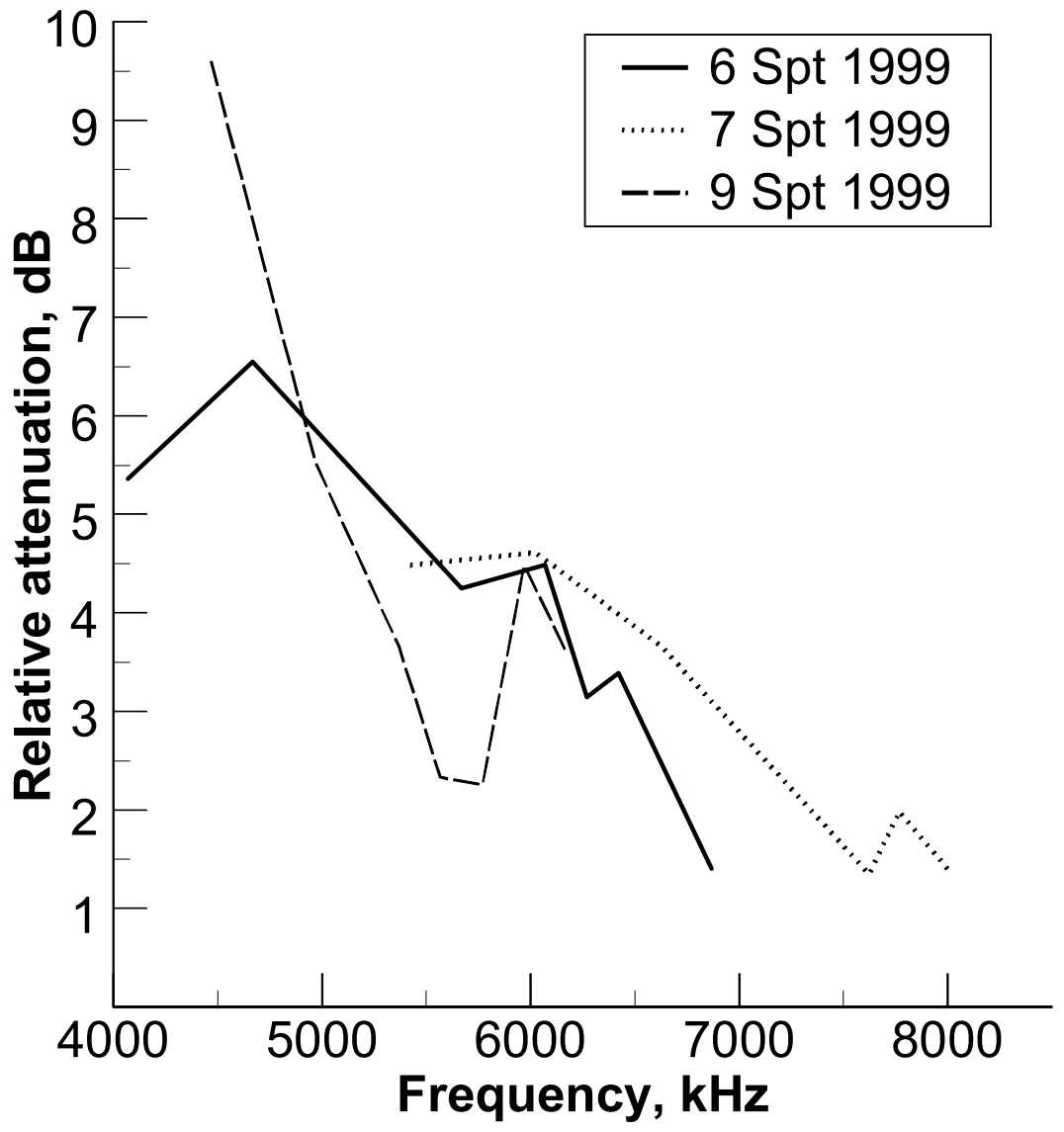

Figure 5.

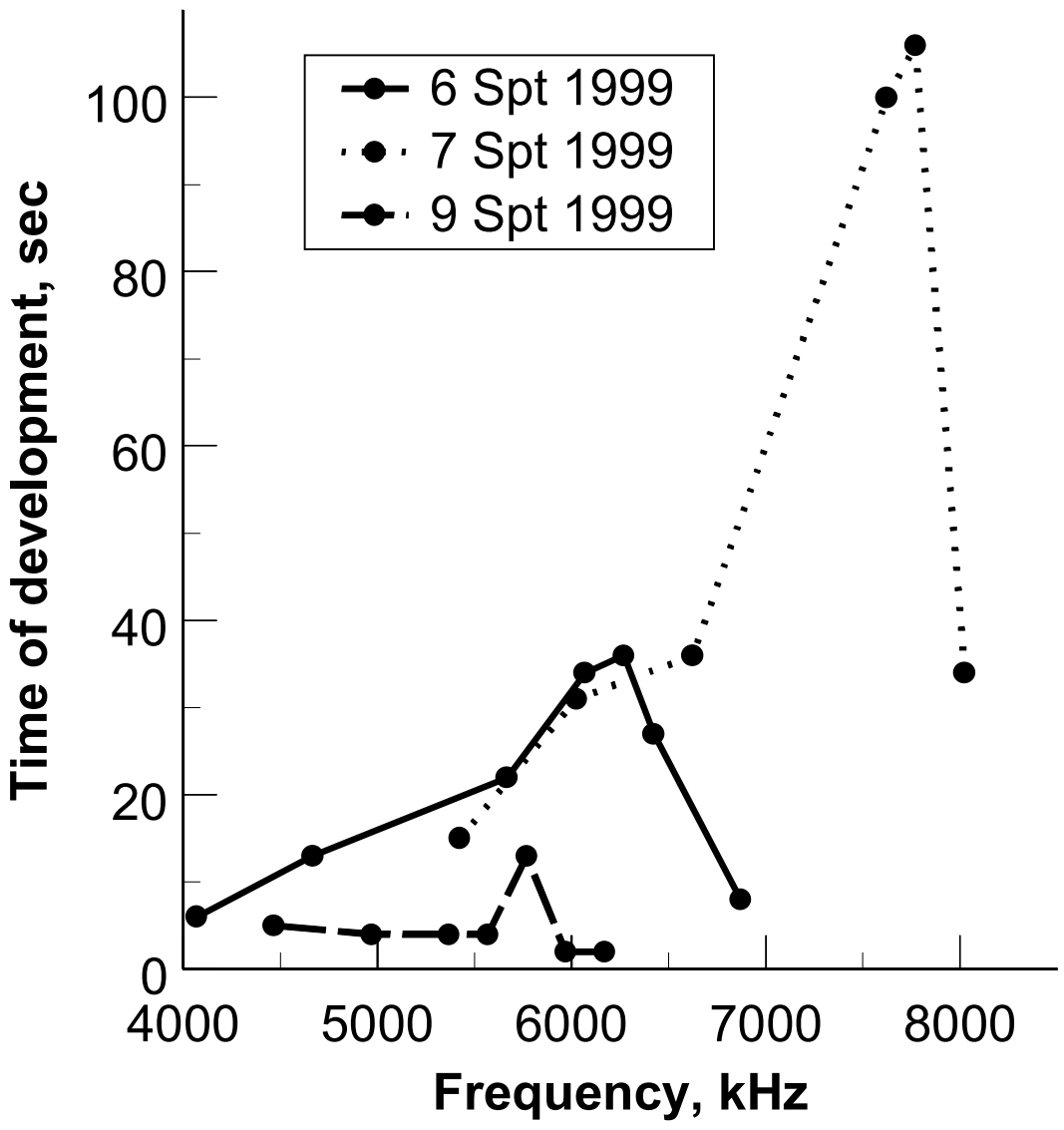

Figure 6.

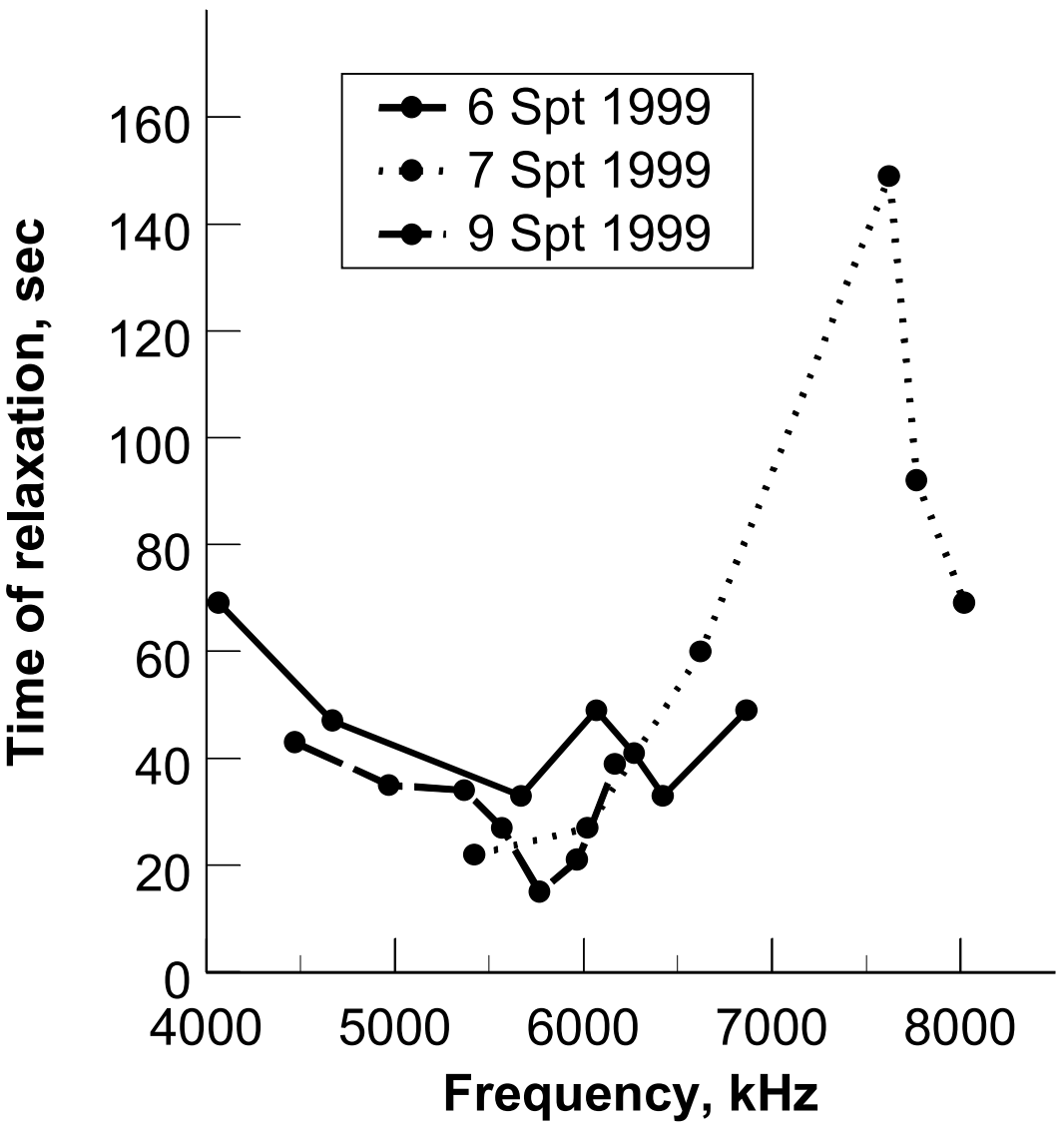

Figure 7.

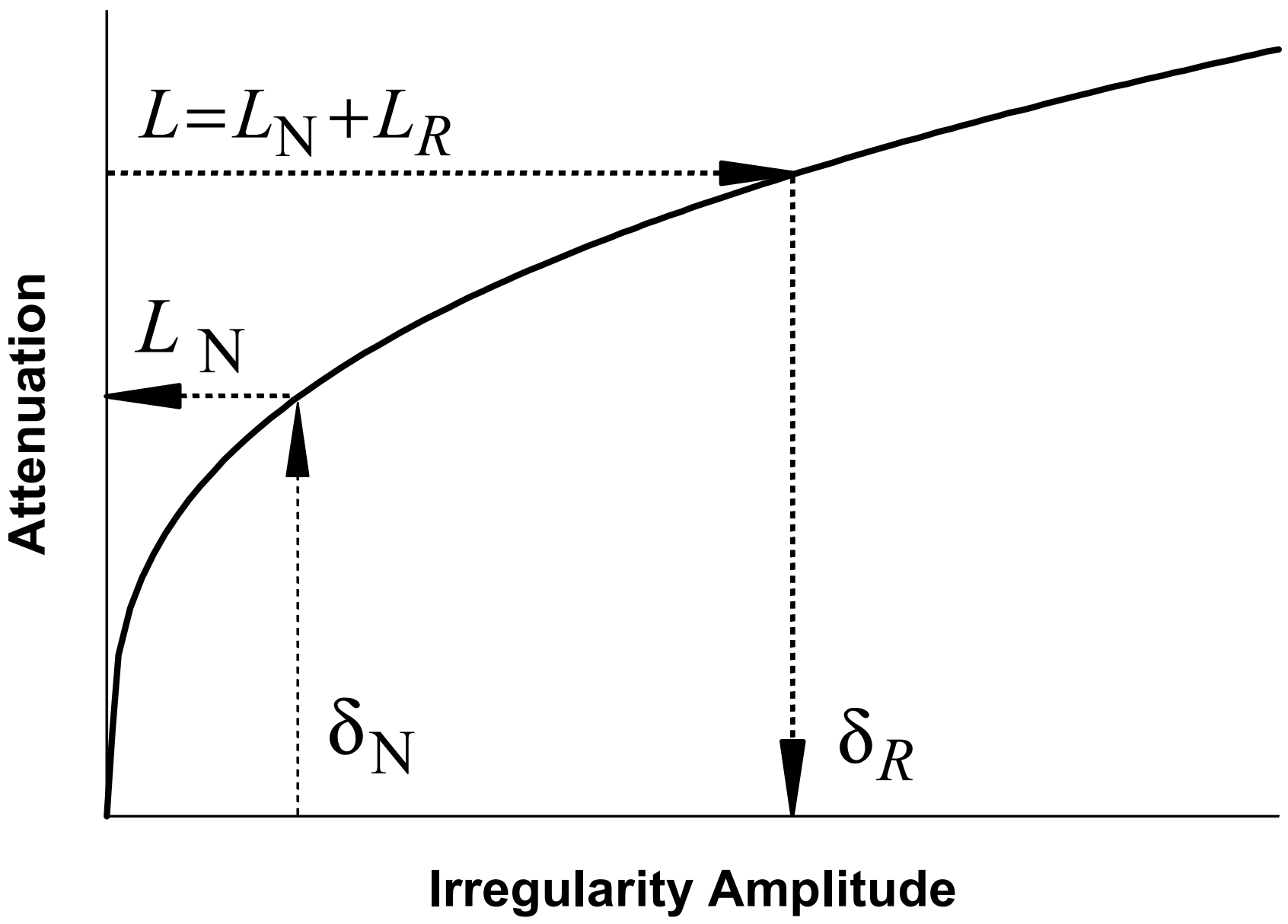

Figure 8.

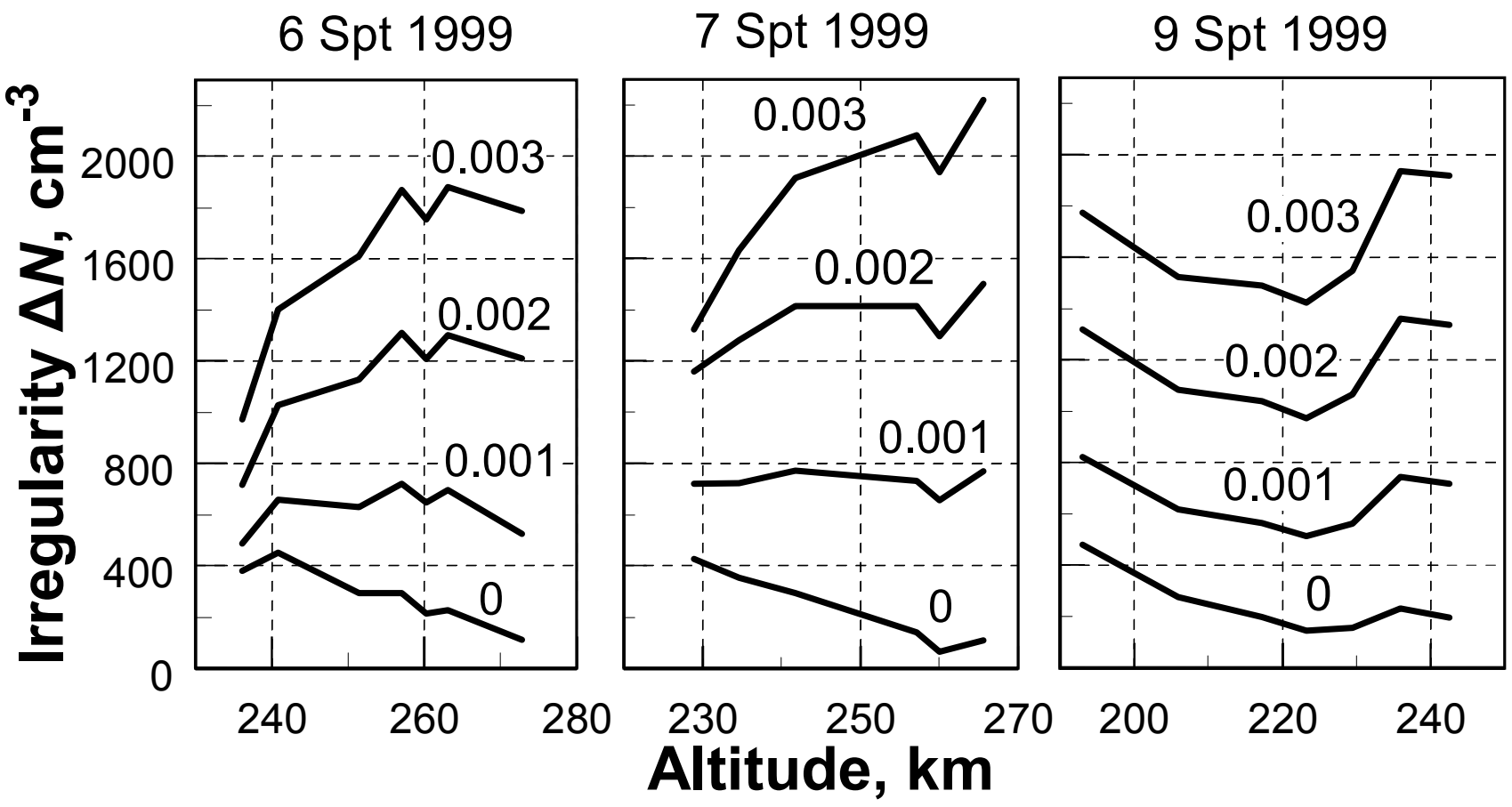

Figure 9.